\documentclass[
 aps, pra,
 amsmath,amssymb,
 12pt,
 final,
tightenlines,
 nofootinbib,
 superscriptaddress,
 ]
{revtex4}

\UseRawInputEncoding
\usepackage[english]{babel}
\usepackage{graphicx}
\def\degr{\hbox{$^\circ$}}
\begin{document}

\title{Long-term optical spectroscopy of B[e] star CI\,Cam in a quiet state }

\author{Valentina G.~Klochkova}
\email{Valentina.R11@yandex.ru}
\affiliation{Special  Astrophysical Observatory,  Nizhnij Arkhyz, 369167 Russia}
\author{Anatoly S.~Miroshnichenko}
\affiliation{University of North Carolina at Greensboro, PO Box 26170, Greensboro, NC 27402-6170, USA}
\affiliation{Fesenkov Astrophysical Institute, Observatory 23, 050023, Almaty, Kazakhstan}

\author{Vladimir~E.~Panchuk}
\affiliation{Special  Astrophysical Observatory,  Nizhnij Arkhyz, 369167 Russia}
\sloppypar
\vspace{2mm}
\noindent

\begin{abstract} 
High-resolution optical spectra of the B[e] star CI\,Cam were obtained on arbitrary
dates 2002--2023 using  the echelle spectrograph NES of the 6-m BTA telescope.
The temporal variability of the powerful emissions of H$\alpha$ and He\,I profiles is found.
For two-peaked emissions with ``rectangular'' profiles, the intensity ratio of blue-shifted
and red-shifted peaks is $V/R \ge 1$, except one date. A decrease in the intensity
of all double-peaked emissions with ``rectangular'' profiles was revealed as they moved
away in time from the 1998 outburst. The average radial velocity for emissions of this type
for all observational dates varies in the range  $(-50.8 \div -55.7)\pm 0.2$\,km/s.
The half-amplitude of the change (standard deviation) is equal to $\Delta$Vr=2.5\,km/s.
The velocity for single-peaked ion emissions (Si\,III, Al\,III, Fe\,III) differs little
from the values of Vr(emis-d), but the measurement accuracy for these emissions is 
worse:  the average error for different dates ranges from 0.4 to 1.3\,km/s. The systemic
velocity is assumed to be Vsys=$-55.4\pm 0.6$\,km/s according to the stable position
of the forbidden emission [N\,II]\,5754\,\AA{}. The position of single-peaked emissions
[O\,III]\,4959 and 5007\,\AA{} is also stable: Vr([O\,III])=$-54.2\pm 0.4$\,km/s.
Forbidden emissions [O\,I]\,5577, 6300, 6363\,\AA{}, [CaII]\,7291 and 7324\,\AA{} are
absent from the spectra. Appearence of the emission near 4686\,\AA{} is an infrequent 
event, its intensity rarely exceeds the noise level. Only a wide asymmetric 
emission with an intensity of about 16\% above the local continuum was registered 
in the spectrum for March~9,~2015. Questions arise about the use of this emission to
estimate the period of variability of the star and about localization of this feature
in the CI\,Cam system. The photospheric absorptions of N\,II, S\,II, and Fe\,III with
a variable position are identified.  
\end{abstract}

\maketitle

\section{Introduction}

The hot star CI\,Cam has long been known as a star with an optical spectrum rich in emissions,
it is designated MWC\,84 in the catalog [1]. Particular interest in this object arose after
the registration of a powerful outburst in its system in April~1998, which was observed in
all spectral ranges from X-ray to radio [2]. After this event, the star was classified as
an ultraluminous X-ray source (for more details, see [3]) and is studied intensively
in all wavelength ranges, the SIMBAD database contains over 250 publications. The main
features of the optical spectrum of CI\,Cam intense emissions of HI, He\,I, as well as emissions
of Fe\,II, [Fe\,II] with specific ``rectangular'' profiles, were identified already in early
articles [4, 5], whose authors classified the star as a supergiant with the B[e] phenomenon.
Let us recall that the first interpretation of emission ``rectangular'' profiles for stars
with constant loss of matter was given by Beals [6], using the example of recorded emission
bands with flat tops in the spectra of Wolf--Rayet stars and Novae. By now, there is a
widespread idea that emissions in the spectra of stars with the B[e] phenomenon are formed
in a structured circumstellar medium in the form of a dense disk and/or arcs. Zickgraph [7]
summarized the fundamental features of the optical spectra of B[e] stars and carried out 
theoretical  modeling of the profiles of specific spectral features. 
The presence of forbidden lines of metals in the optical spectrum caused by their 
formation within a low-density  circumstellar medium close to the star, is considered 
as the main feature of stars with the B[e] phenomenon.

Just a few days after the 1998 outburst CI\,Cam spectroscopy in the IR range revealed
powerful emissions of HI, He, Fe and their subsequent changes [8]. The results of the
first observations in various wavelength ranges stimulated a number of observational
campaigns, the results of which were later summarized in numerous original articles and
reviews. Let us note the most important publications for understanding the features
of the optical spectrum of CI\,Cam (and related objects): these are [3, 7--9].
Important for understanding the features of stars with the B[e] phenomenon is the work [10].
The authors of this work, having obtained high-resolution spectra for a sample of stars,
studied the variability of emissions and came to the conclusion that specific emissions
arise in a set of envelopes structures. To study the long-term behavior of the optical
spectrum, CI\,Cam has no analogues with high spectral resolution observations made
two weeks after the 1998 outburst at the 2.7-m telescope of the McDonald Observatory,
which allowed the authors  [11] identify the components of the spectrum in detail,
estimate their parameters and indicate the areas of formation of various types of
emissions.

Several groups of astronomers have studied the time behavior of photometric and
spectral parameters of CI\,Cam. The principal results were obtained through
spectral monitoring carried out by the authors~[12], which tracked changes in
the profiles of the main emissions in the optical spectrum during 2001--2005,
in particular, a transition of the H$\alpha$ profile was recorded from double-peak
type to single-peaked type. These authors highlighted the significant error in
determining equivalent widths of spectral features due to continuum level
uncertainty. An important resultthe detection of an inclined dust disk -- was obtained
by interferometry in the near-infrared range [13].
Goranskij~et~al. [14], using an extensive collection of spectra obtained with  some
spectrographs of the 6-m BTA telescope, providing different spectral resolutions,
made a number of significant conclusions. In particular, the study of the radial
velocity Vr  pattern along lines of various natures (He\,II\,4686\AA{}, [N\,II]\,5754\,\AA{}
and  Fe\,II) revealed a special behavior of the forbidden emission [N\,II]\,5754\,\AA{},
which allowed the authors to conclude that this line is formed in a region more
distant from the star compared to that for the Fe\,II emissions.

To date, the main ideas about the structure and behavior of CI\,Cam can be briefly 
summarized as follows: CI\,Cam is a B[e] phenomenon object, which includes a massive 
star of early spectral class B and a low-mass companion of an as yet unknown nature;
several years after the 1998 outburst, the type of emissions in the optical spectrum
changed and their intensity decreased significantly; the complex structure of the
circumstellar medium was formed due to fast polar wind and a slow equatorial outflow;
there are no absorption features in the optical spectrum, with the exception of
interstellar absortions; there are no reliable estimates of the distance, luminosity
and orbital parameters; for example, there is no generally accepted value for
the orbital period.

The need for numerous high-quality spectral data is obvious, so publications often 
end with a call for such observations. However, the large distance
of the CI\,Cam (parallax $\pi$=0.21\,mas according to Gaia~EDR3 [15]) hinders the 
regular high-resolution spectral monitoring needed to generate a comprehensive
collection of spectra. As M.~Kraus  correctly noted in her review [9], performing  
such monitoring requires a lot of patience and a lot of time on large 
telescopes. So far, only a few studies have been published with results
based on the analysis of high-resolution spectra obtained at individual 
points in time, for example,  [3, 5]. In an effort to increase  time observations 
interval, the authors are forced to combine measurement data obtained from 
observations with high and low spectral resolution [14].

The lack of observational data of the required quality and volume served 
as an argument for us to begin long-term spectroscopy to search for 
variability in the profiles of spectral features and the radial velocity 
pattern in time. This task requires repeated observations with high 
spectral resolution over a wide range of wavelengths. In this article, 
we present the results of the first stage of work. Section~2 of this 
article briefly describes the methods of observation and data analysis.
Section~3 presents the results obtained, and sections~4 and 5 discuss 
the results in comparison with previously published ones, as well as 
summarize the main conclusions

\section{Echelle spectroscopy at BTA}  

Our first observations of CI\,Cam at the BTA were made using a moderate 
resolution spectrograph  (R=12000) PFES~[16] on the night of December~3, 1998. 
This spectrum, containing only single-peaked emissions due to insufficient 
resolution, showed us the need and possibility of obtaining CI\,Cam 
spectra with a much higher resolution on the BTA. All subsequent CI\,Cam
spectra were obtained by us at  high resolution  in 2002--2023 with 
echelle spectrograph NES~[17], stationary located at the Nasmyth focus 
of the 6-m BTA telescope. The dates of observations  are given in Table\,1.

In recent years, the spectrograph has been equipped with a CCD detector 
with the number of elements 4608$\times$2048, element size  
0.0135$\times$0.0135\,mm; readout  noise 1.8e$^-$. Wavelength range 
recorded in one exposure $\Delta\lambda 470\div$778\,nm. In 2001--2011
we used a CCD with a number of elements 2048$\times$2048. To reduce
flux losses at the entrance slit, the NES spectrograph is equipped with 
an imahe slicer. Using a slicer, each spectral order is repeated three times. 
The spectral resolution of the NES spectrograph  R=$\lambda/\Delta\lambda\ge 60\,000$.
In our CI\,Cam spectra, the signal-to-noise ratio (S/N) varies by several
orders of magnitude from the continuum to the peaks of strong emissions.

\small{
\begin{table}[htbp]
\medskip
\caption{Results of heliocentric radial  velocity measurements in  the
  spectra of CI\,Cam  for different types of fetures. The number of
  details used in averaging the velocity  is indicated in parentheses}
\begin{tabular}{ c| c|  c|  c| c|c  }
\hline
Date & $\Delta\lambda$ &\multicolumn{4}{c}{\small Vr, km/s} \\
\cline{3-6}
   JD   & nm & Emis-d &Emis-s &[OIII]&  DIBs   \\
\hline
1&2&3&4&5& 6 \\
\hline
04.02.2002 &462-607   &$-47.4\pm1.3$&  &$-51.2$    & $-7.8$   \\  [-10pt]
2452310.34 &          & (43)        &  && (2) \\ [-10pt]
19.11.2002 &456-599   &$-49.6\pm0.2$&$-47.4\pm0.8$ & -48.0& $-7.4\pm0.7$ \\ [-10pt]
2452598.16 &          & (46)    &(25)  &   &  (8)  \\ [-10pt]
21.12.2002 &452--599  &$-51.7\pm0.2$&$-51.5\pm0.5$ &$-55.2$& \\  [-10pt]
2452630.44 &          & (28)        &  (12)        &  & \\[-10pt]
18.11.2005 &528--670  &$-53.8\pm0.2$&$-54.1\pm0.5$ &  &$-7.8\pm0.7$ \\ [-10pt]
2453693.40 &          &  (42)       &    (8)       &  & (9)\\ [-10pt]
20.11.2005 &528--670  &$-54.6\pm0.4$&$-54.8\pm1.0$ &  &$-6.0\pm0.5$\\ [-10pt]
2453694.51 &          &  (21)       &   (11)       &  & (15) \\ [-10pt]
15.01.2006 &456--601  &$-54.4\pm0.2$&$-54.9\pm0.2$ &$-51.6$& \\ [-10pt]
2453751.20 &          &    (39)     &   (8)        &  & \\  [-10pt]
06.12.2006 &447--593  &$-55.1\pm0.2$&$-53.1\pm1.3$ &$-54.7$&$-7.2\pm1.0$ \\ [-10pt]
2454076.36 &          &  (61)       &   (14)       &  &  (6)   \\ [-10pt]
07.12.2006 &447--593  &$-55.7\pm0.2$&$-53.3\pm0.4$ &$-55.8$& $-7.8\pm0.9$   \\ [-10pt]
2454077.22 &          &  (61)       &   (26)       &  & (5) \\ [-10pt]
08.12.2006 &447--593  &$-55.6\pm0.2$&$-56.3\pm0.5$ &$-54.8$ & $-6.8\pm0.6$    \\ [-10pt]
2454078.35 &          &  (59)       &   (19)       &  & (7) \\ [-10pt]
10.12.2006 &447--593  &$-55.6\pm0.2$&$-56.2\pm0.5$ &$-56.7$& $-7.0\pm1.0$   \\ [-10pt]
2454080.35 &          &  (59)       &   (22)       &  &  (7) \\ [-10pt]
18.11.2008 &447--593  &$-55.6\pm0.2$&$-55.7\pm0.6$ &$-56.3$ & \\  [-10pt]
2454788.29 &          &  (21)       &   (9)        &  &  \\ [-10pt]
13.01.2011 &520--668  &$-54.0\pm0.2$&$-52.4\pm0.4$ &  & $-6.6\pm0.7$ \\ [-10pt]
2455575.26 &          &  (35)       &  (12)        &  & (7)\\ [-10pt]
03.09.2015 &395--698  &$-54.4\pm0.2$& $-54.0\pm0.4$&$-55.1$&$-7.3\pm0.6$  \\ [-10pt]
2457269.47 &          & (88)        & (35)         &  &(17)  \\ [-10pt]
08.12.2019 &470--778  &$-51.6\pm0.2$&$-51.9\pm0.7$ &$-43.4$& $-7.6\pm0.3$   \\ [-10pt]
2458826.45 &          &   (46)      &   (32)       &  &   (19)      \\  [-10pt]
21.10.2021 &470--778  &$-51.5\pm0.3$&$-51.3\pm0.5$ &$-43.0$ &$-6.2\pm0.8$ \\ [-10pt]
2459509.33 &          & (49) & (23) &   &(17)  \\ [-10pt]
05.07.2022 &470--778  &$-50.8\pm0.2$& $-50.0\pm0.2$&-48.9  &$-6.2\pm0.5$  \\ [-10pt]
2459890.37 &          & (56)        &  (32)          & (5)          &(18)  \\ [-10pt]
09.02.2023 &470--778  &$-51.1\pm0.3$& $-53.6\pm0.5$&-48.4  &$-7.1\pm0.5$  \\ [-10pt]
2459985.28 &          &     (60)    &   (12)       &   &   (10)  \\
\hline
\end{tabular}
\label{velocity}
\end{table}
}

Extraction of one-dimensional data from two-dimensional echelle frames 
is performed using the ECHELLE context of the MIDAS package modified 
taking into account the geometry of the echelle frame.
All details of the procedure are described in [18]. Traces of cosmic particles
were removed using a standard technique -- by median averaging a pair of
spectra obtained sequentially.
A Th--Ar lamp was used to calibrate the wavelengths.
All subsequent steps in the processing of one-dimensional spectra were 
performed using the modern version of the DECH20t package [19]. Systematic
error in heliocentric radial velocity measurements are based on a set of 
telluric  features and interstellar lines of the Na\,I doublet, it does not
exceed 0.25\,km/s for one line; measurement error for broad
absorptions does not exceed 0.5\,km/s. For average velocity values in Table\,1,
errors are 0.6--2.4\,km/s depending on the type of the  measured lines. 
We identified features in the CI\,Cam spectra using lists of lines
from papers based on spectroscopy at the BTA+NES of related stars with
the B[e] phenomenon [20--22].
In addition, for several spectral details we used data
from the VALD database (see [23] and references therein).

\section{Main results}

\subsection{Features of the CI\,Cam spectrum and their variability}

The optical spectrum of CI\,Cam is populated by emissions of different types: 
emissions of metal ions with two-peaked ``rectangular'' profiles (hereafter -- Emis-d),
complicated not entirely symmetrical emissions of H\,I, He\,I, weak emissions of multiply
ionized metals (Al\,III, Fe\,III, Si\,III, hereafter -- Emis-s), emission components
of the oxygen  triplet OI\,7773\,\AA{}. All of them are formed under different
physical conditions  in the circumstellar environment and have different widths.
For this reason,  we examined separately the time behavior of the profiles of
these types of emissions and radial velocity from measurements of  their positions.

In Figs.\,1 and 2, these main features of the star\`{s} spectrum are well 
illustrated.  According to Beals classification [6], the H$\alpha$ profile  belongs
to type~II -- its single-peaked emission that does not contain absorption components.
Figure~1 shows the H$\alpha$ profiles for several observational dates. The position 
of the dashed vertical line in this and subsequent figures corresponds to 
 the accepted value  Vsys=$-55.4\pm 0.6$\,km/s for the [N\,II] line 5754\,\AA{}.
Let us note the features detected in the  H$\alpha$ profile: systematic decrease 
in the intensity of the profile top; lack of wind components; presence, in
addition to a wide pedestal ($\Delta$Vr$=\pm400$\,km/s), a slight increase in
the intensity of the long-wavelength wing; its variability may be caused by 
the inhomogeneitie of the  circumstellar medium.

\begin{figure}[ht!]
\includegraphics[angle=0,width=0.8\textwidth,bb=40 40 720 530,clip]{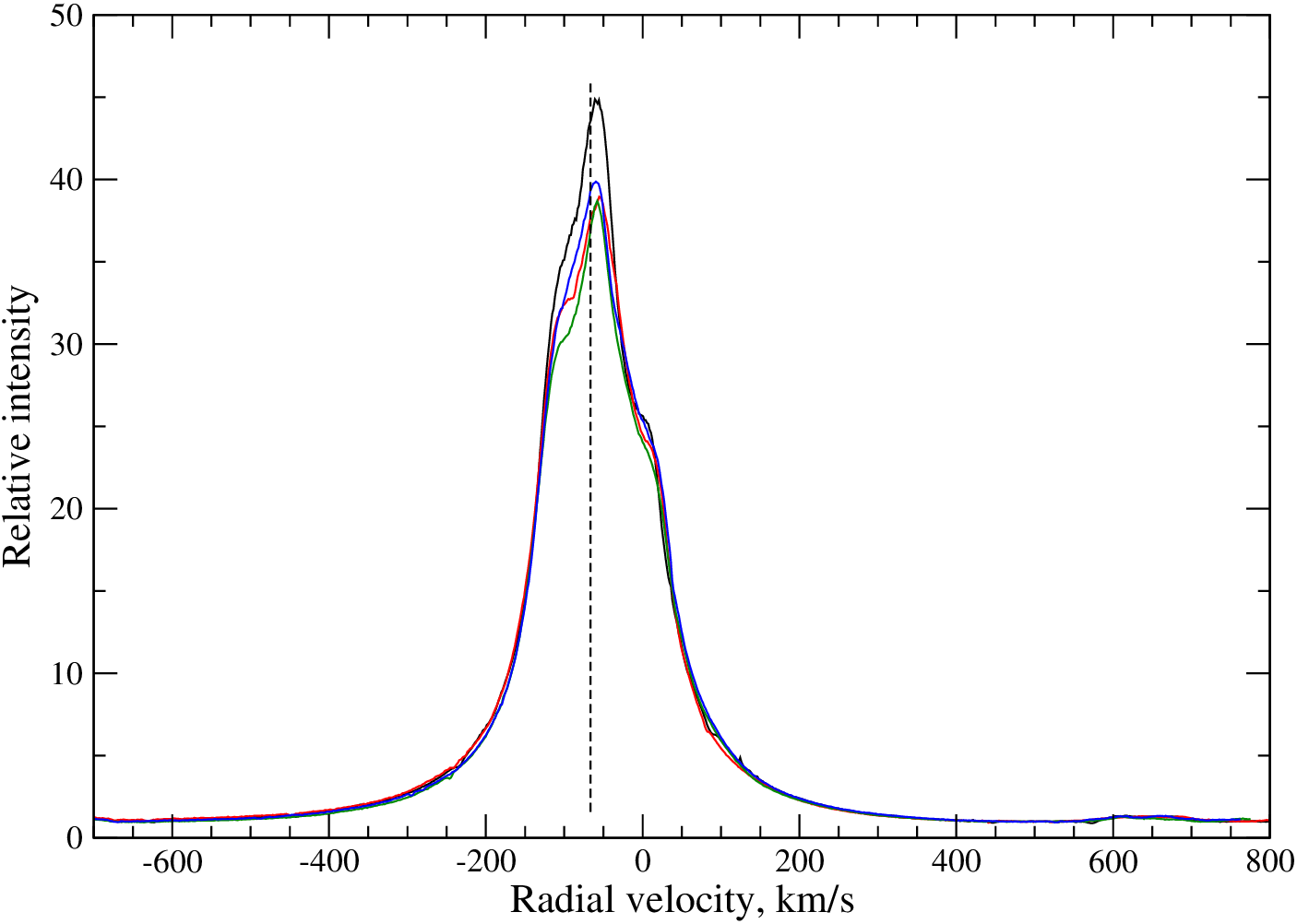}
\caption{H$\alpha$ Profiles in CI\,Cam spectra obtained on different dates:
November~18,~2005 (orange line), September~3,~2015 (black),
October~21,~2021 (red), July~5,~2022 (green), February~9,~2023 (blue).
Here and in subsequent figures, the position of the dashed
vertical corresponds to the accepted value Vsys=$-55.4\pm0.6$\,km/s for the
[N\,II] line 5754\,\AA{}, and the continuum level is taken as 1. }
\label{Halpha}
\end{figure}

\begin{figure}[ht!]
\includegraphics[angle=0,width=0.6\textwidth,bb=10 80 560 680,clip]{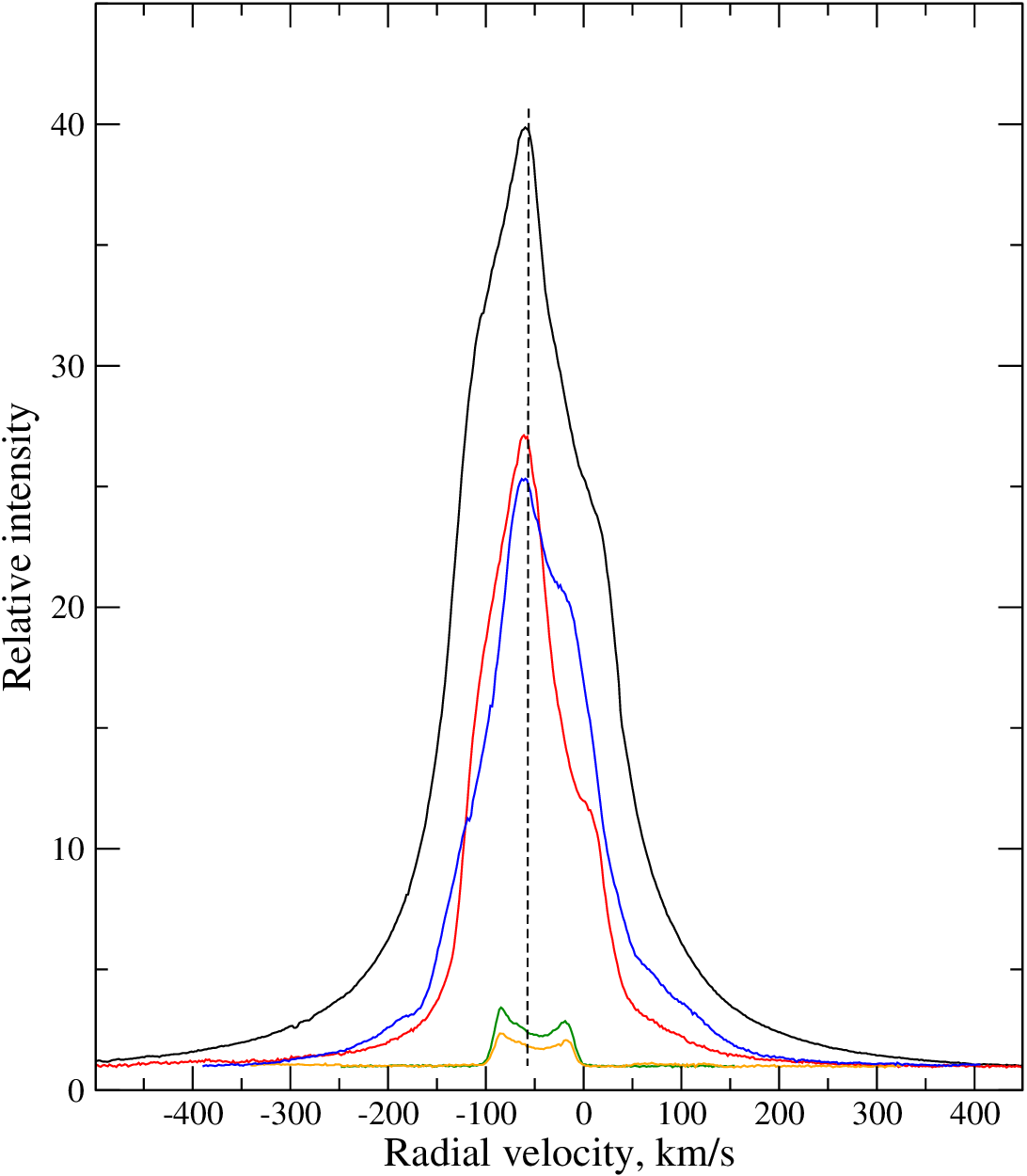}
\caption{Profiles of specific emissions in the spectrum on February~9,~2023:
H$\alpha$ (black line), H$\beta$ (red),  He\,I~5876\,\AA{} (blue), Fe\,II~5534\,\AA{}
(green), Fe\,II~5425\,\AA{} (orange)}
\label{Lines}
\end{figure}

The profiles of double-peaked emissions (Fe\,II, [Fe\,II], etc.) have almost 
vertical slopes and concave peaks. In the case of a spherically symmetric and 
radially expanding wind, profiles with a flat top are formed in an optically 
thin medium. The observed shape of the vertices in the CI\,Cam spectrum indicates the
structure of  emitting layer. According to [3], the concavity of the profile, i.e., 
a decrease in emission, is due to less radiation at lower velocities. Even in the 
case of symmetry of the circumstellar medium, the tops of the profiles will not 
be flat even  if there is additional absorption in the disk.

\begin{figure}[ht!]
\vbox{
\includegraphics[angle=0,width=0.4\textwidth,bb=30 80 560 680,clip]{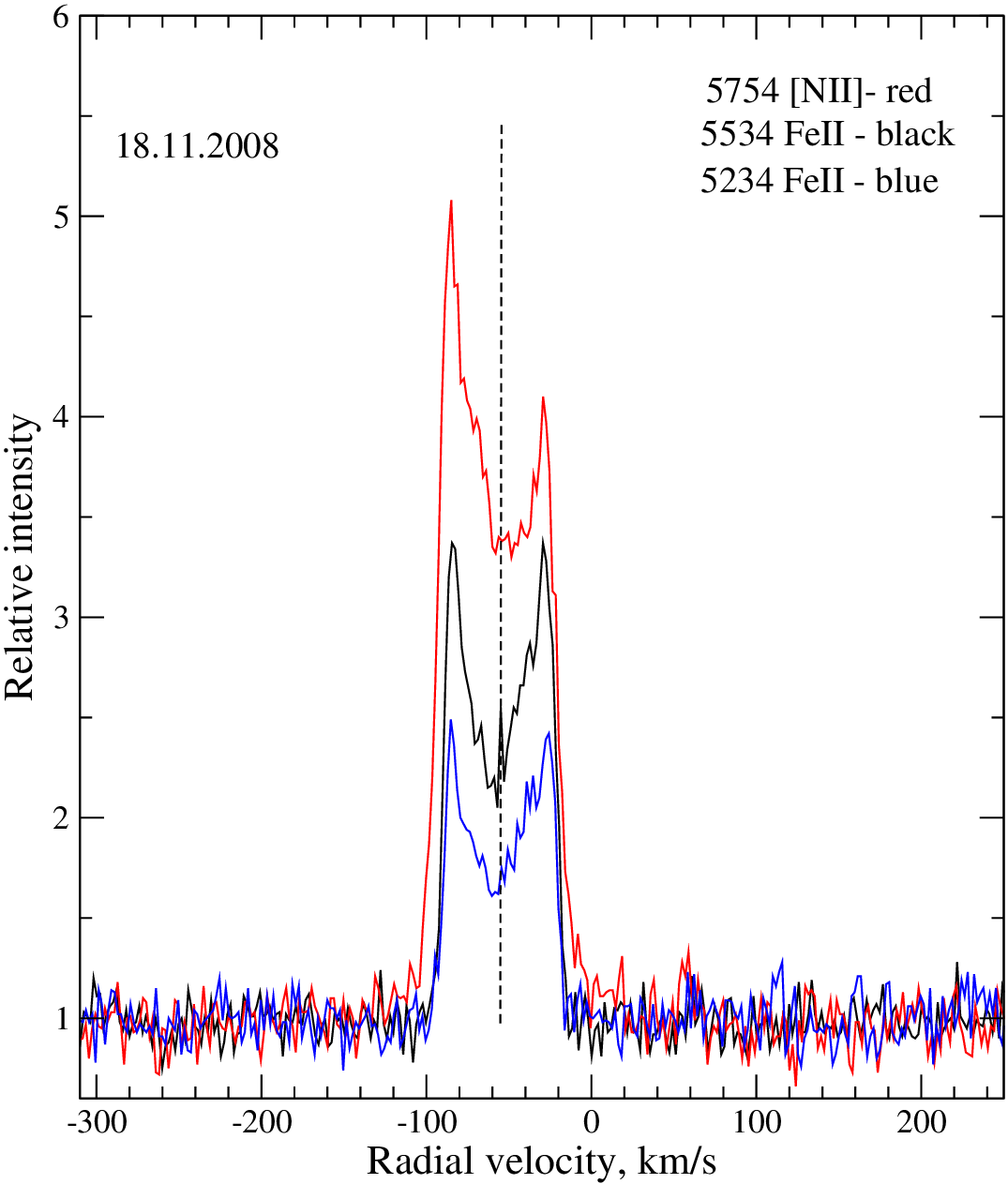} 
\includegraphics[angle=0,width=0.4\textwidth,bb=30 80 560 680,clip]{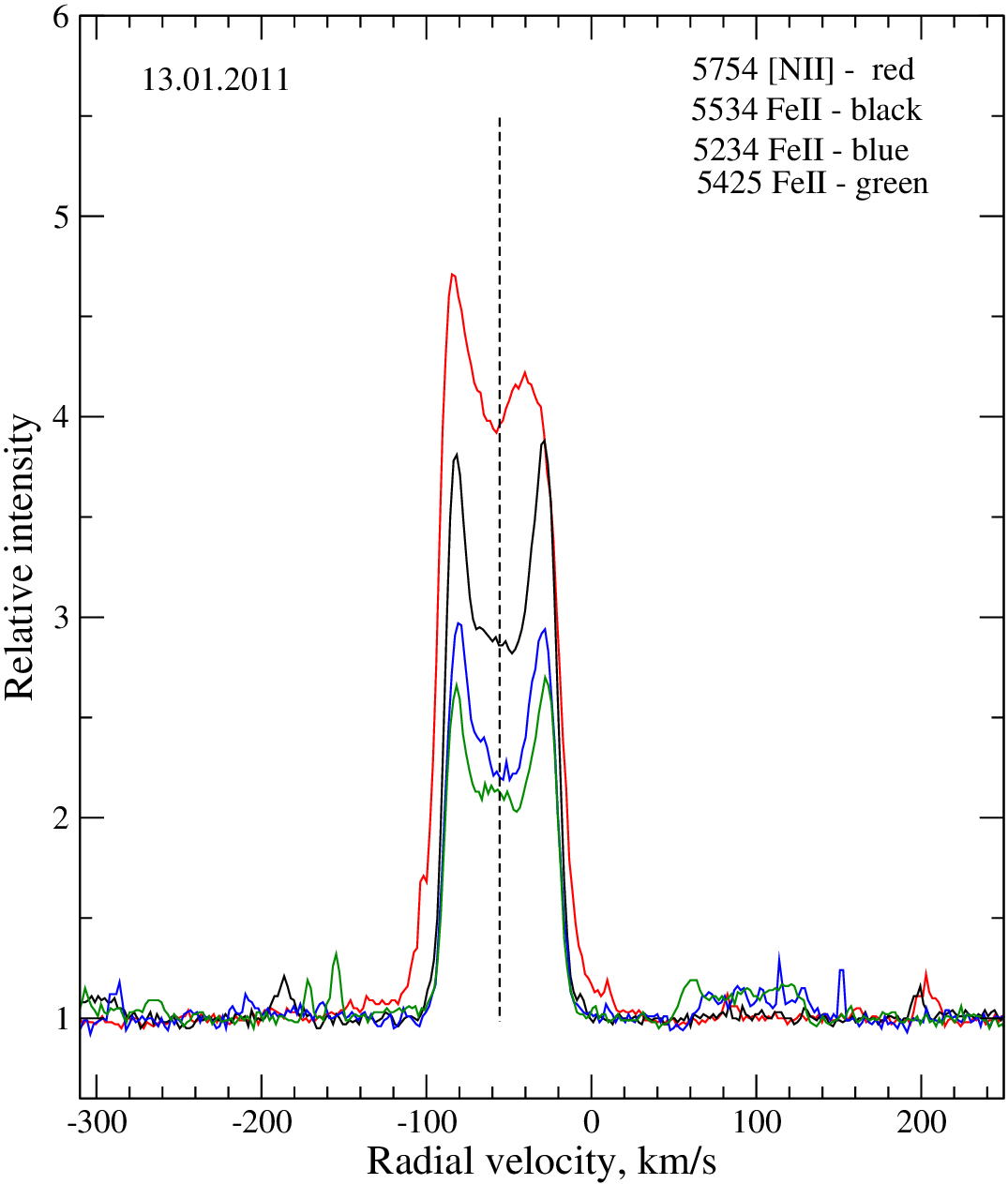} 
\includegraphics[angle=0,width=0.4\textwidth,bb=30 80 560 680,clip]{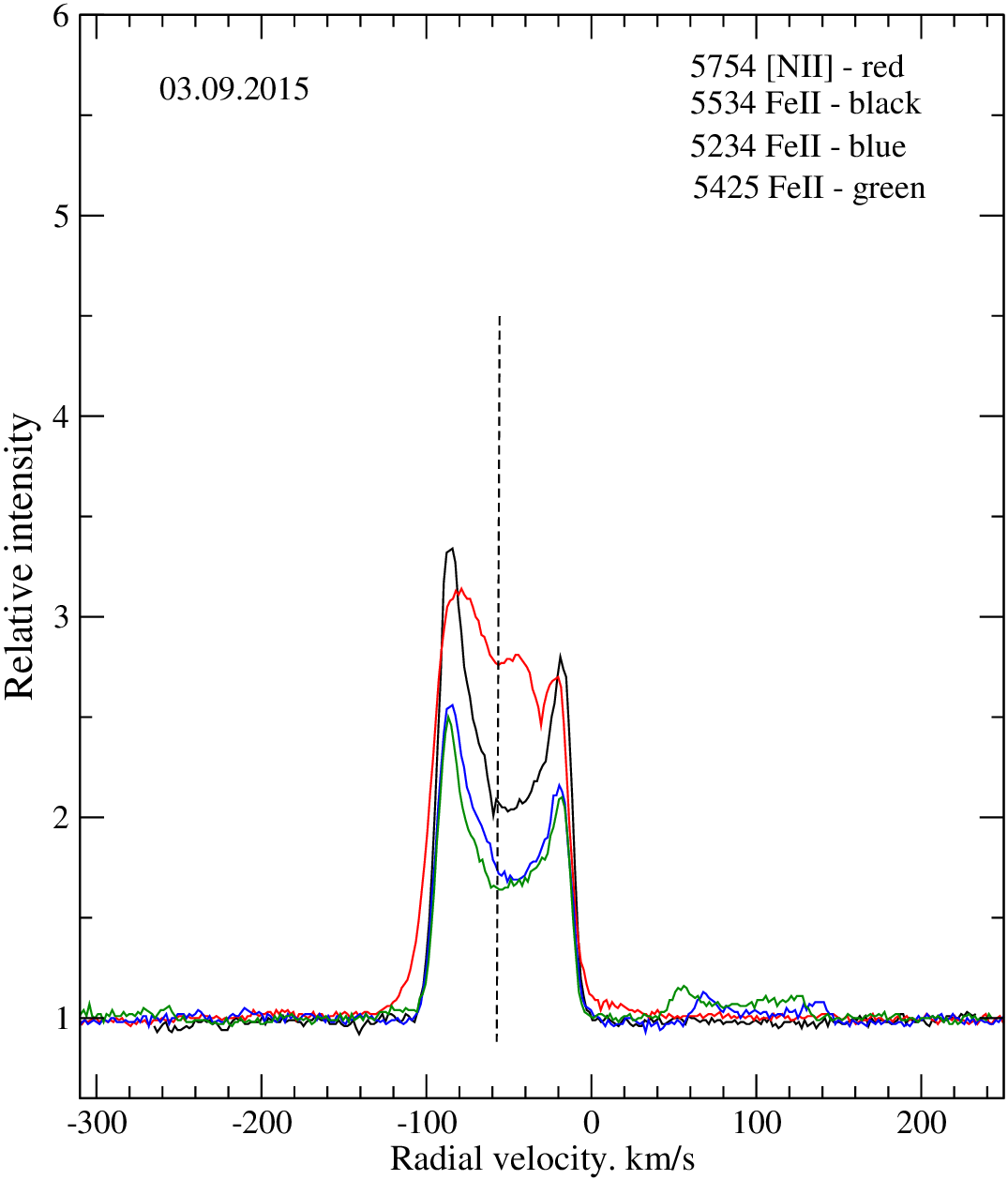} 
\includegraphics[angle=0,width=0.4\textwidth,bb=30 80 560 680,clip]{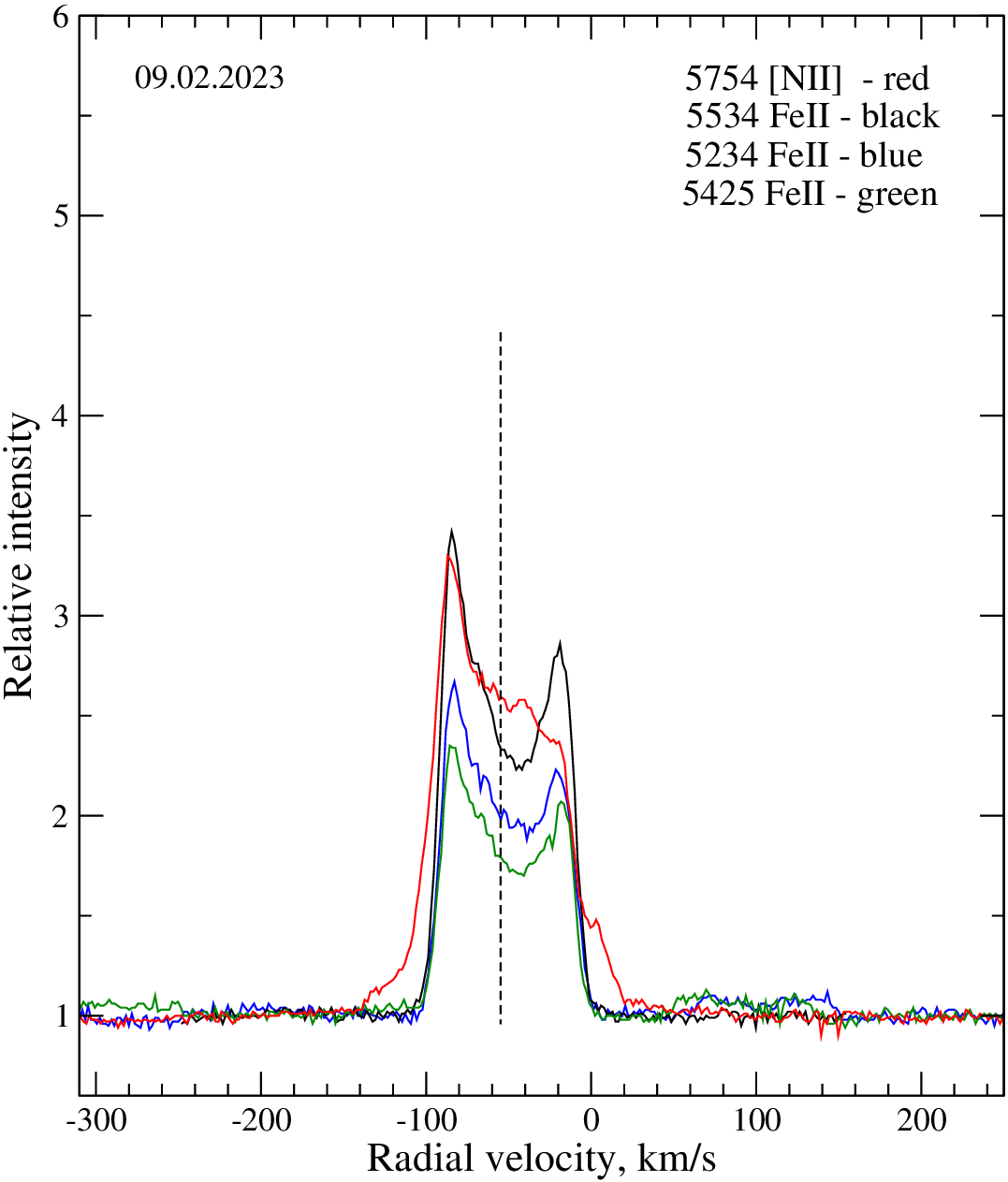} 
}
\caption{Profiles of selected lines in the spectra obtained on November~18,~2008,
January~13, 2011,  September 3, 2015, and February~9, 2023: [N\,II]~5754\,\AA{}
(red line), Fe\,II~5534\,\AA{} (black), Fe\,II~5234\,\AA{} (blue),
Fe\,II~5425\,\AA{} (green).}
\label{Emis-var}
\end{figure}

The variability of Fe\,II and [Fe\,II] emissions with ``rectangular'' profiles 
 and concave peaks is illustrated  in Fig.\,3. This specific shape of Fe\,II and 
[Fe\,II] emissions  persists during monitoring, however, the intensity
of the emission peaks in the spectra obtained on different dates systematically 
changes. Noteworthy  the right   top panel, which shows a rarely
observed ratio of peak intensities: the intensity of the long-wavelength 
peak for all emissions (except for the forbidden emission [N\,II]\,5754\,\AA{})
is higher than the  short-wavelength one.

On the left panel of Fig.\,4, the profiles of the [N\,II]\,5754\,\AA{} line
in two early spectra of CI\,Cam (02/04/2002 and 11/19/2002) were compared with
profiles obtained from 2008 to 2022. Here we see a significant difference between 
the early profiles of 2002 relative to all subsequent ones, which have the same
the position of the short-wavelength wings and the intensity changes insignificantly. 
At the same time, in the earliest profile of 02/04/2002 we see an almost
twofold increase in intensity, and a shift in position, a shift of the short-wave 
wing by about 10\,km/s toward longer wavelengths. Already 8.5 months later, in 
November~2002, the intensity of this profile decreased significantly,
and its short-wavelength wing approached the stationary position of the later 
profiles. The ratio $V/R$ exceeds 1 for all observational epochs. The right
panel of this figure shows the Fe\,II~5534\,\AA{} line profile for the
same dates for comparison. For this line of low excitation in the spectra of 
2002 we also see a shift of the short-wavelength wing by about 10\,km/s  red,
but the behavior of the $\rm V/R$ value is different: the intensity changes 
insignificantly, in the spectra of 2002 the ratio $V/R \ge 1$, later
$V/R < 1$.

\begin{figure}[ht!]
\hbox{
\includegraphics[angle=0,width=0.45\textwidth,bb=30 70 560 680,clip]{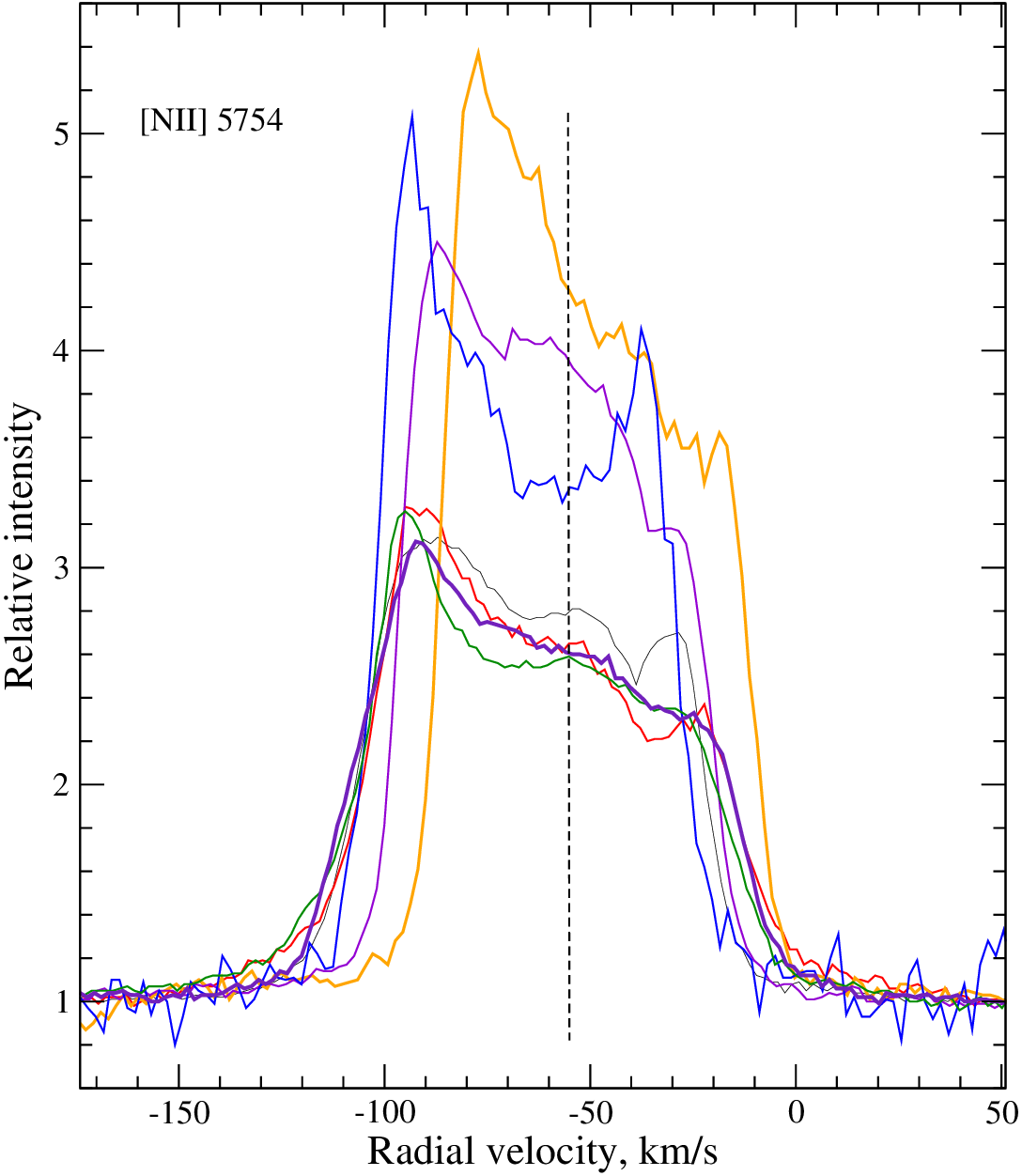}  
\includegraphics[angle=0,width=0.45\textwidth,bb=30 70 560 680,clip]{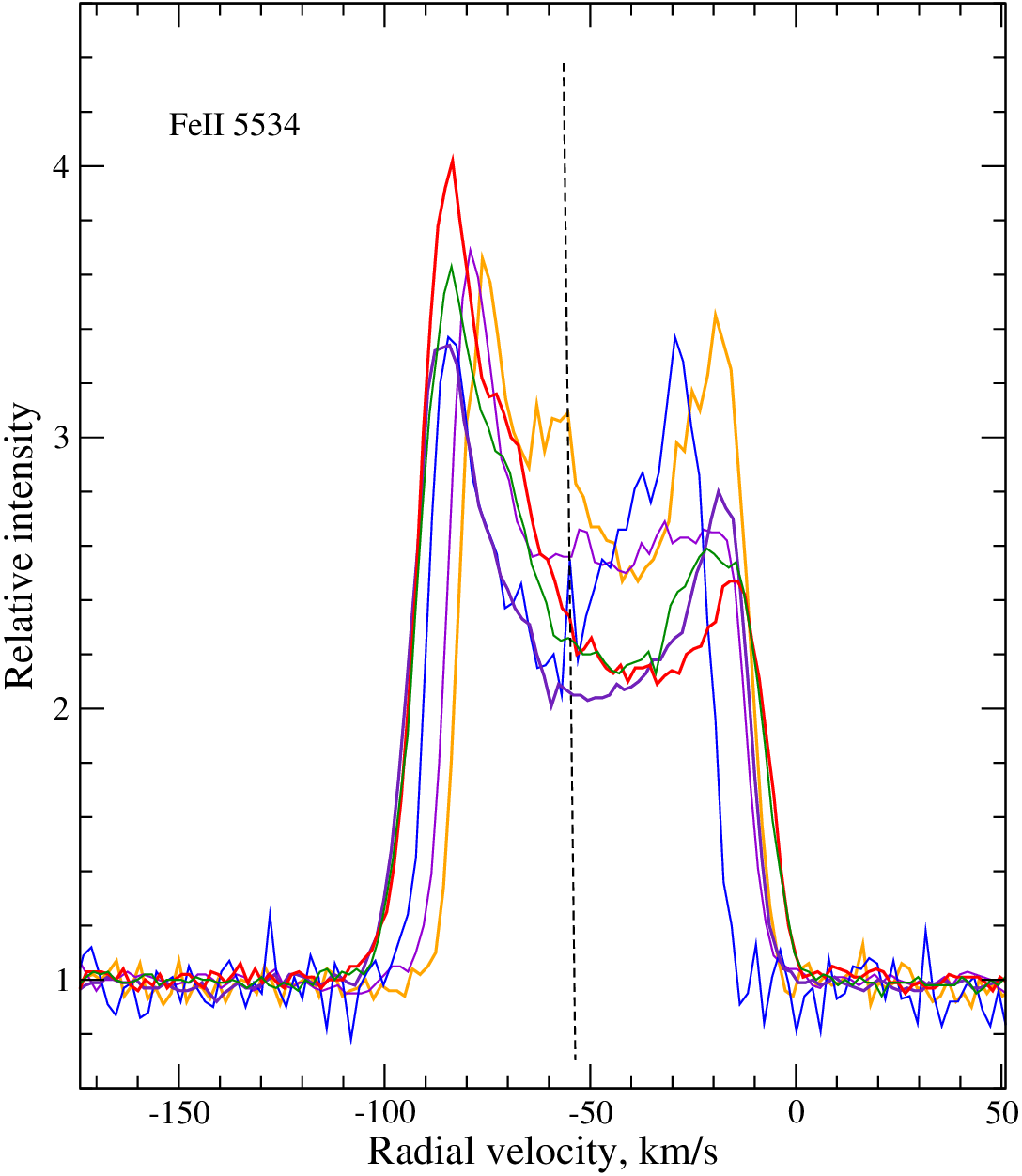}  
}
\caption{Emission profiles of [N\,II]~5754\,\AA{} (left) and Fe\,II~5534\,\AA{}
(right) in spectra of different dates: February~4,~2002 (orange line),
November~19,~2002 (purple), November~18,~2008 (blue), September~3,~2015 (black),
December~8,~2019 (thick black line), October~21,~2021 (red), July~5,~2022 (green).}
\label{5754_5534}
\end{figure}

Publications of the results of optical spectroscopy of CI\,Cam, as a rule, 
contain an indication of the presence of high excitation emissions in the spectra:
He\,II~4686\,\AA{}, forbidden emissions [OI]~6300 and 6363\,\AA{}, [O\,III]~4959 
and 5007\,\AA{} (see, for example, works [11, 12]). The [O\,I]~6300 and 6363\,\AA{} 
emission spectra are absent in our spectra, but the [O\,III]~4959 and 5007\,\AA{} 
lines are constantly observed. The identification of these weak lines of high 
excitation [O\,III]~4959 and 5007\,\AA{} does not cause problems, and the
positions of both lines in the spectra,  Vr(4959)=$-52.4\pm 1.6$ and  
Vr(5007)=$-50.6\pm1.4$\,km/s (averaged from  13 spectra),  are in good agreement 
and change little over time.

\begin{figure}[ht!]
\includegraphics[angle=0,width=0.6\textwidth,bb=15 60 550 730,clip]{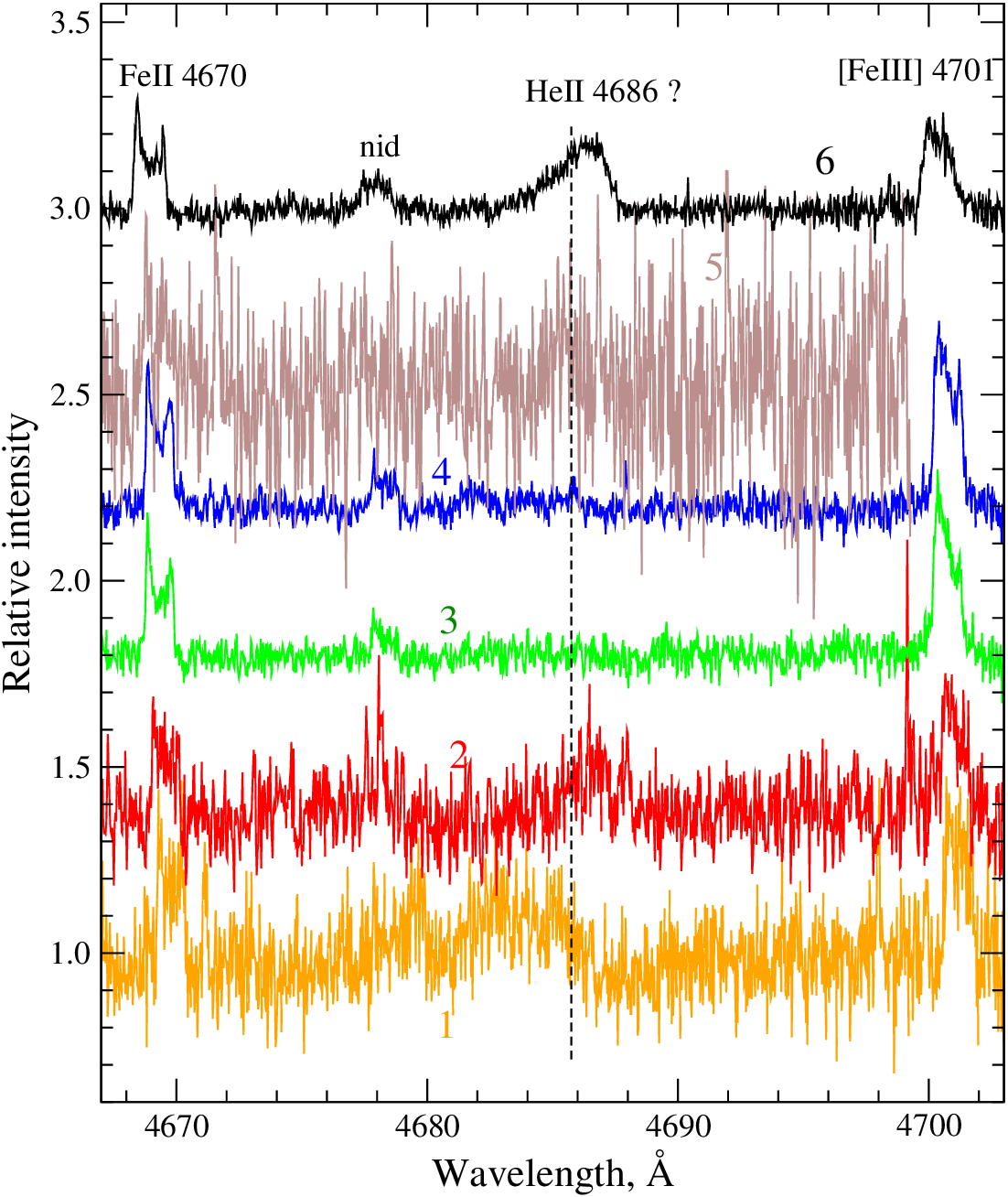}   
\caption{Fragments of CI\,Cam spectra in the wavelength range 4665--4703\,\AA{},
obtained on various dates: February~4,~2002 (orange line), January~15,~2006 (red),
December~6,~2006 (green), December~7,~2006 (blue), November~18,~2008 (brown),
September~3,~2015 (black line). The vertical position corresponds to a wavelength
of 4686\,\AA{}. A weak emission at a wavelength around 4678\,\AA{} has not been
identified.}
\label{4686}
\end{figure}

The situation with the behavior of the He\,II~4686\,\AA{} emission in 
high-resolution spectra remains less certain. Eight of our spectra listed in 
Table\,1 contain the wavelength range in which the presence of the 4686\,\AA{} 
line can be expected. In Fig.\,5, fragments for 6 observational 
dates are presented. In spectra ``1'', ``3'', ``4'' and ``5'' this line is 
absent; in the spectrum ``2'' the emission intensity is low (about $<5\%$ above the
local continuum). In our earliest spectrum, obtained on December~3,~1998, with 
moderate resolution and S/N$>100$  with the PFES spectrograph, there is no 
emission at 4686\,\AA{}. And only in the spectrum, obtained on September~3,~2015, 
did a broad emission appear with an equivalent width of W$_\lambda=0.4$\,\AA{}. 
If we assume that this is the He\,II emission 4686\,\AA{}, we obtain the width 
of its profile in the range of radial velocities from  $-100$ to $+160$\,km/s, 
which does not agree with the general picture of radial velocities in the system 
(see line profiles in Fig.\,2) and raises the question of the localization of
this emission in the CI\,Cam system.

The He\,II~4686\,\AA{} emission line may not appear in all orbital phases due, 
for example, to screening by dense regions of the shell of the main star of the system.
In addition, it is weak and may not be visible against the background noise in spectra 
with low S/N ratious. Based on our observations, we can recognize the appearance 
of the He\,II~4686\,\AA{}  emission in the spectra of CI Cam as a rare event,
which indicates the difficulty of using this emission to estimate the period of thr 
spectral variability of the star and raises the problem of localizing the emission in
the CI\,Cam system. To solve this problem, there is an obvious need to obtain a 
significantly larger volume of high-quality spectral data. There is currently not
enough high-resolution spectral data to support the use of this line for this purpose.

We also note that in the spectrum obtained on September~3,~2015, in which the 
emission  was recorded near $\lambda$=4686\,\AA{}, an additional feature is 
observed -- a reduced intensity of emissions with ``rectangular'' profiles,
which is well illustrated in the set of panels in Fig.\,4.

\begin{figure}[ht!]
\includegraphics[angle=0,width=0.49\textwidth,bb=40 40 725 525,clip]{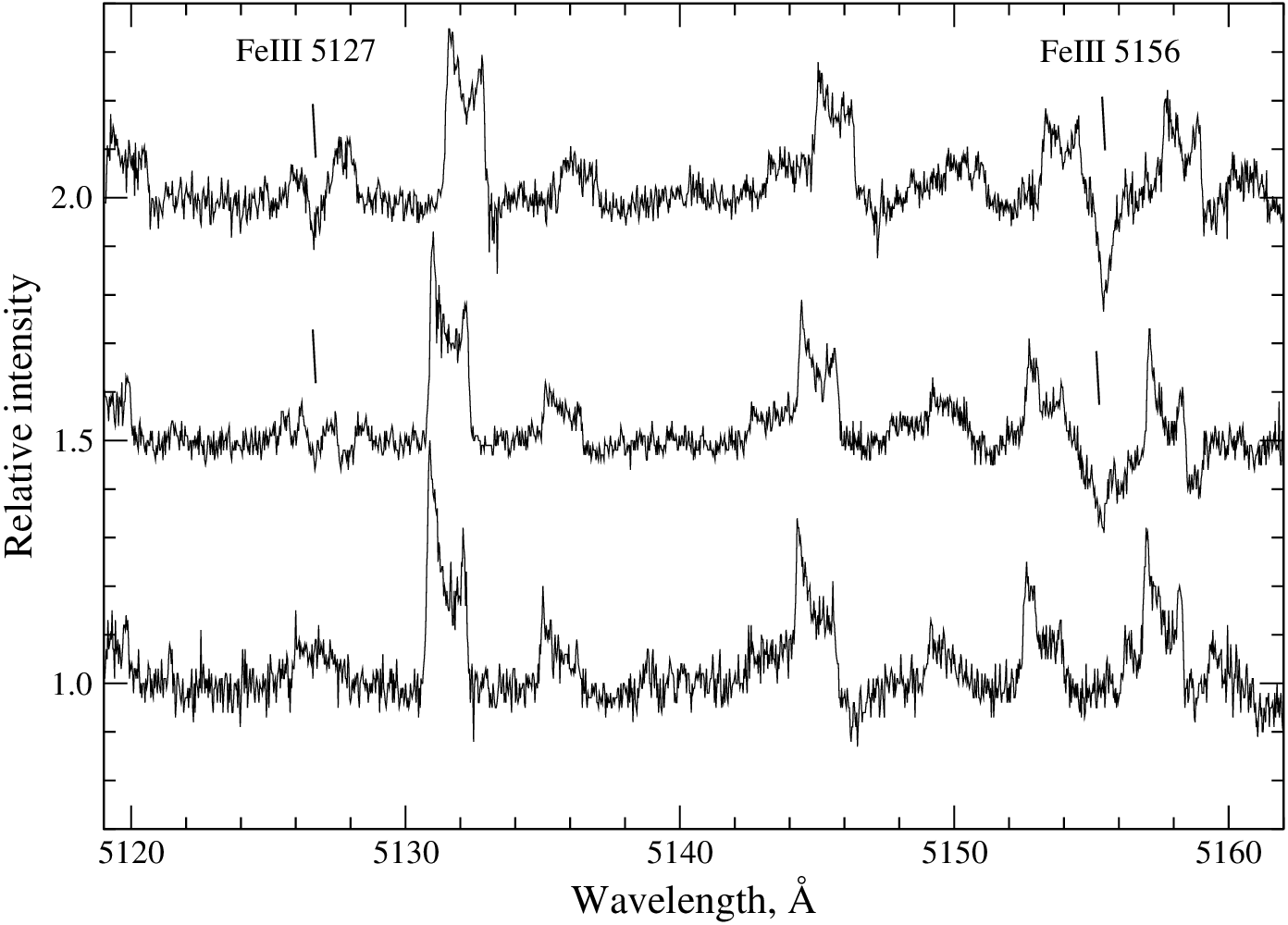}
\includegraphics[angle=0,width=0.49\textwidth,bb=40 40 725 525,clip]{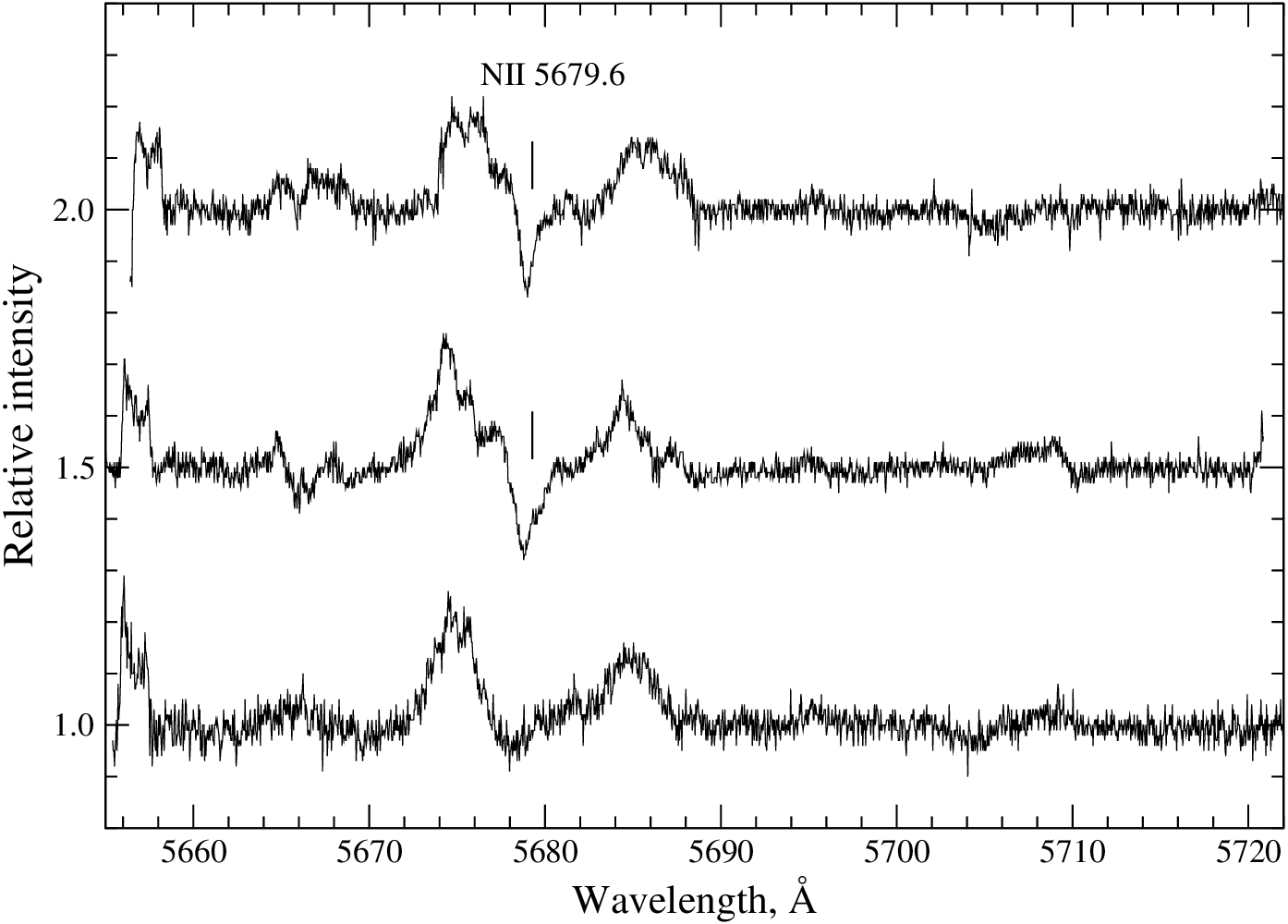}
\caption{Fragments with absortions in the CI\,Cam spectra, from top to bottom:
February,~9,~2023, July~5,~2022, October~21,~2021. The two upper fragments are
shifted along the ordinate axis by 0.5 and 1.0, respectively}
\label{Absorb}
\end{figure}

\subsection{Search for absorptions in CI\,Cam spectra}

A close relative of CI\,Cam is the supergiant MWC\,17 with the B[e] phenomenon, 
the features of which were studied in detail using several spectra of the BTA\,+\,NES 
by the authors [21]. The star MWC\,17, identified  with the IR source IRAS\,01441+6026, 
is one of the hottest stars with  the B[e] phenomenon. In its spectra taken in 2005--2006 
numerous permitted and forbidden emissions, as well as interstellar Na\,I lines and diffuse 
interstellar bands (DIBs), have been identified. A comparison of the new data
obtained with earlier measurements allowed us to conclude that there is no 
significant variability in spectral details. The pattern of radial velocities was 
studied using lines of various dates, which allowed the authors to accept the 
velocity of stable forbidden emissions as the systemic velocity, Vsys=$-47$\,km/s.
The high quality of the spectral material used allowed the authors [21] perform 
detailed identification of spectral structures. A careful search did not lead to 
detection of absorptions formed in the stellar atmosphere in the spectrum of MWC\,17.
Having a sample of high-quality CI\,Cam spectra, we also undertook a search for 
similar absorptions, including a comparison of CI\,Cam spectra obtained on different
dates. In the spectra of two observatioaln dates in 2022 and 2023, we identified 
the absorptions of Fe\,III, N\,II and S\,II, and obtained average velocity with a
good  accuracy, Vr(abs-2022)=$-32.4\pm 0.5$\,km/s (5 lines) and Vr(abs-2023)=$-54.6\pm 1.0$\,km/s
(9 lines), which gives rise to the conclusion about the presence of absorptions
and the variability of their positions. In the spectrum obtained on December~8,~2019, 
we found only 5 absorptions; the average  velocity  for this date was determined
with lower accuracy Vr(abs-2019)=$-55.6\pm 2.0$\,km/s.
The position of three of the found absorptions in two fragments of the spectra
for three observational dates is shown in Fig.\,6. From a comparison of fragments of
the spectra of 2021, 2022 and 2023  we concluded  on  the absence of these absorptions
in the spectrum of 2021. In general, we can talk about the presence of absorptions
at certain moments of observations and the variability of their average position. 
Future spectral observations will serve to refine CI\,Cam\`{s} atmospheric spectrum 
information

\subsection{CI\,Cam system velocity}

\begin{figure}[ht!]
\includegraphics[angle=0,width=0.4\textwidth,bb=40 50 720 530,clip]{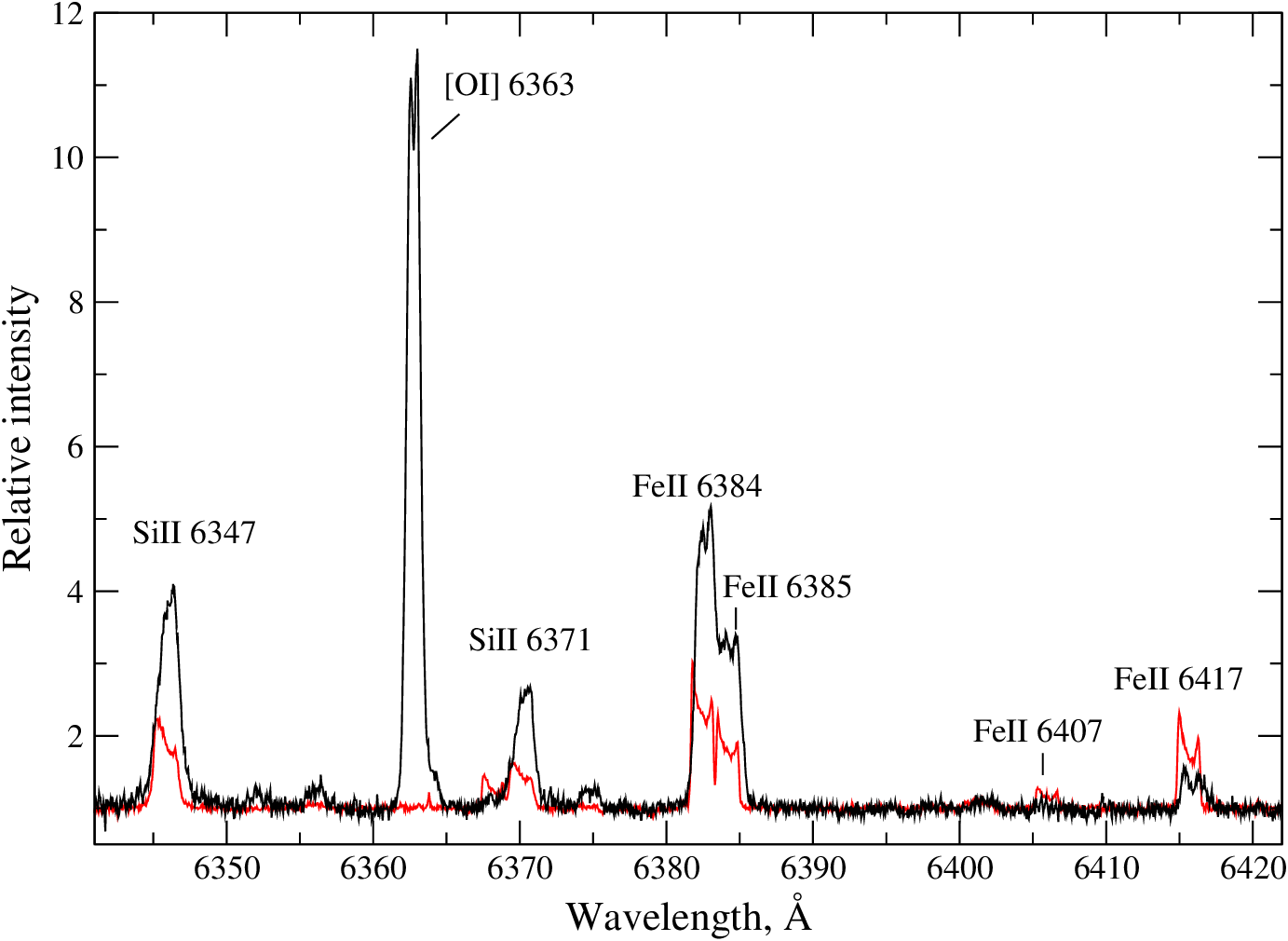}
\includegraphics[angle=0,width=0.4\textwidth,bb=40 50 720 530,clip]{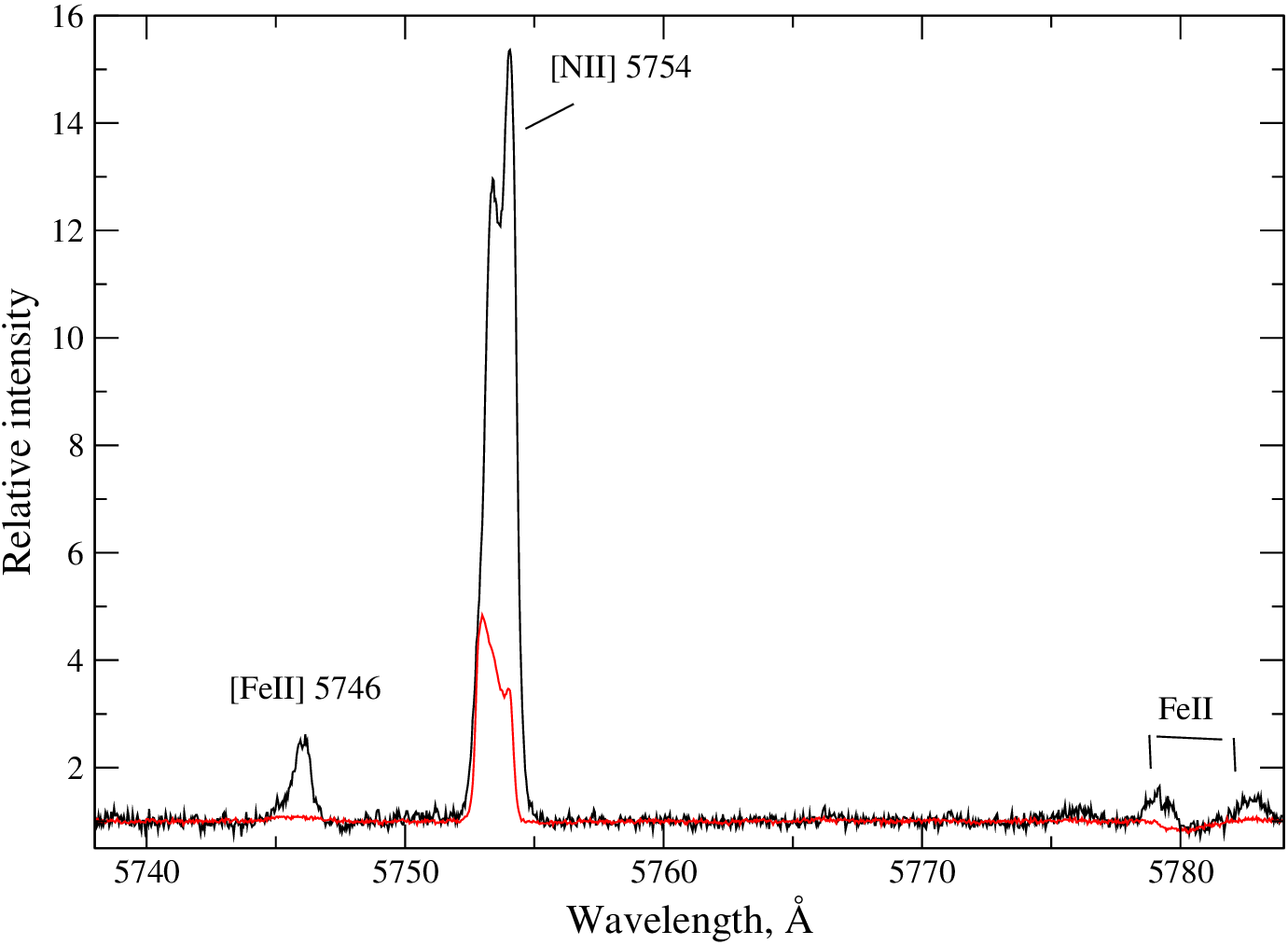}
\caption{Comparison of fragments containing H$\alpha$ and forbidden lines [N\,II],
in the spectra of MWC\,17 (black line) and CI\,Cam (red), obtained on the BTA with
 the NES  spectrograph in December~2005.}
\label{2stars}
\end{figure}

The radial velocity corresponding to the position of the forbidden line 
[N\,II]~5754\,\AA{} is stable throughout all observation nights. The average velocity
for this emission in 13 spectra was  Vr(5754)=$-55.4\pm 0.6$\,km/s and, as mentioned
above, we accepted it as systemic, i.e., Vsys(5754)=$-55.4\pm 0.6$\,km/s. Two other
forbidden [N\,II] lines with wavelengths 6548 and 6584\,\AA{} in the spectrum of CI\,Cam 
are located on the extended (up to 600\,km/s) wings of powerful H$\alpha$ emission, 
and the second of them is also blended by closely spaced C\,II emissions, which somewhat 
reduces the measurement  accuracy.

\subsection{Comparison of the spectra of CI\,Cam and other stars with the B[e] phenomenon}

From a comparison of the spectra of two B[e] supergiants, MWC\,17 and CI\,Cam, 
it follows not only their fundamental similarity, but also differences in details. 
Let us emphasize the difference in the types of emission profiles: in the CI\,Cam 
spectra outside the outburst, the profiles of H$\alpha$, H$\beta$, Fe\,II, [Fe\,II], 
[O\,III]~5007\,\AA{} are single-peaked, but in the spectra of MWC\,17 they are all 
double-peaked, which indicates a difference in the geometry and kinematics of the 
circumstellar medium. H$\alpha$ and [N\,II]  lines in the spectra of both stars are shown 
in Fig.\,7. In addition, the spectra of MWC\,17 contain a number of emissions, 
[OI]~6300, 6363\,\AA{}, that are absent in the spectra of CI\,Cam. On the top panel 
of Fig.\,8, fragments of the spectra of MWC\,17 and CI\,Cam obtained with the NES 
spectrograph in December~2005 are compared. Both fragments contain the same lines, 
but their intensities differ significantly. The most powerful emission in this 
fragment, [OI]~6363\,\AA{}, prevails in the spectrum of MWC\,17 and is completely 
absent in the spectral fragment of CI\,Cam. Of the forbidden oxygen
emissions, only the [O\,I]~5577\,\AA{} emission is constantly present in all our 
CI\,Cam spectra, the profile width and position of which indicate  telluric origin of
this feature. M.\,Kraus [9] notes that forbidden [O\,I] emissions are formed in 
regions of the circumstellar medium with a low electron density.

\begin{figure}[ht!]
\includegraphics[angle=0,width=0.5\textwidth,bb=30 40 570 700,clip]{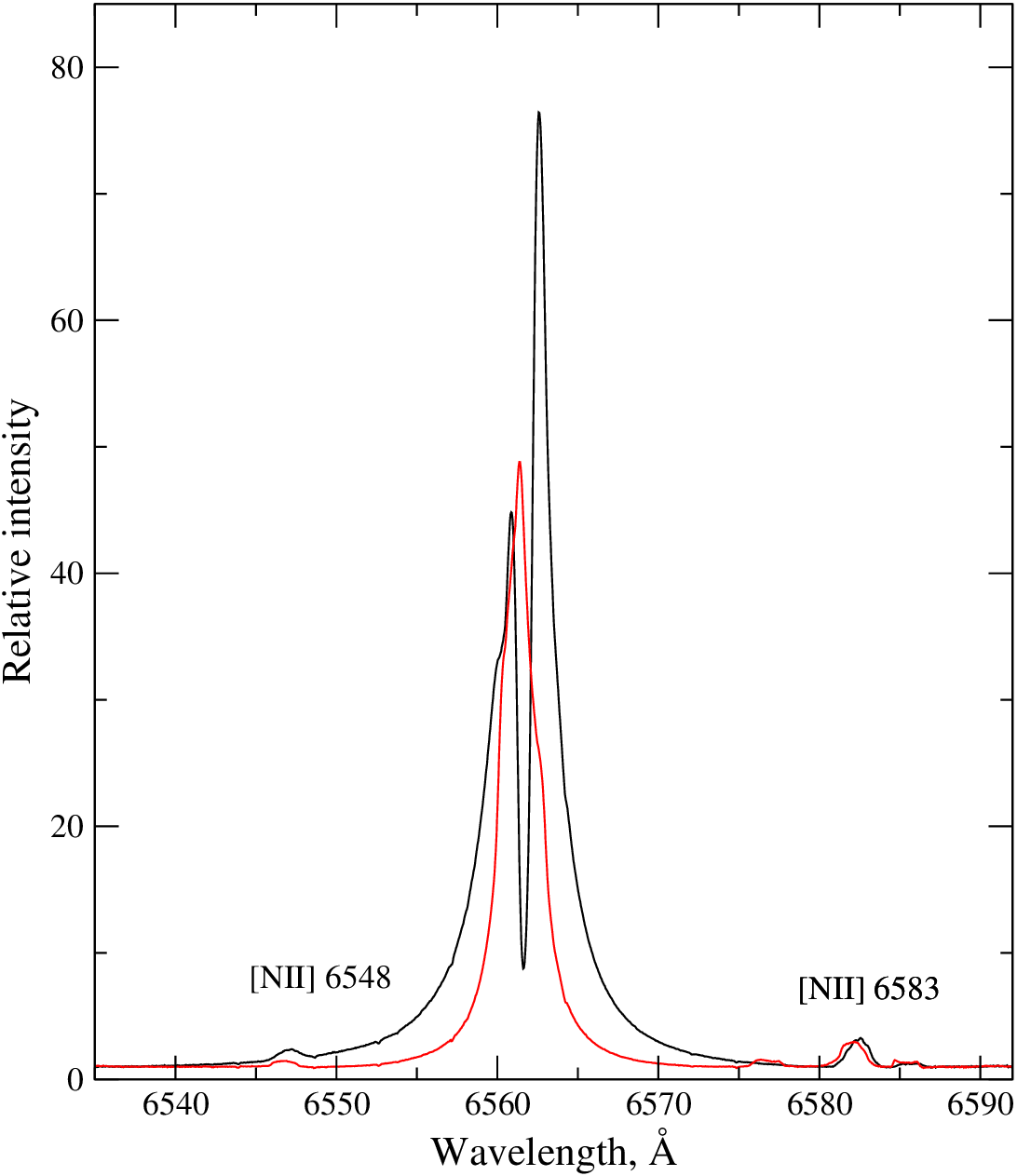}
\caption{Comparison of fragments of the spectrum of MWC\,17 (black line) and CI\,Cam (red),
obtained on the BTA with the NES spectrograph in December 2005.}
\label{2stars-Halpha}
\end{figure}

\begin{figure}[ht!]
\includegraphics[angle=0,width=0.45\textwidth,bb=10 80 560 780,clip]{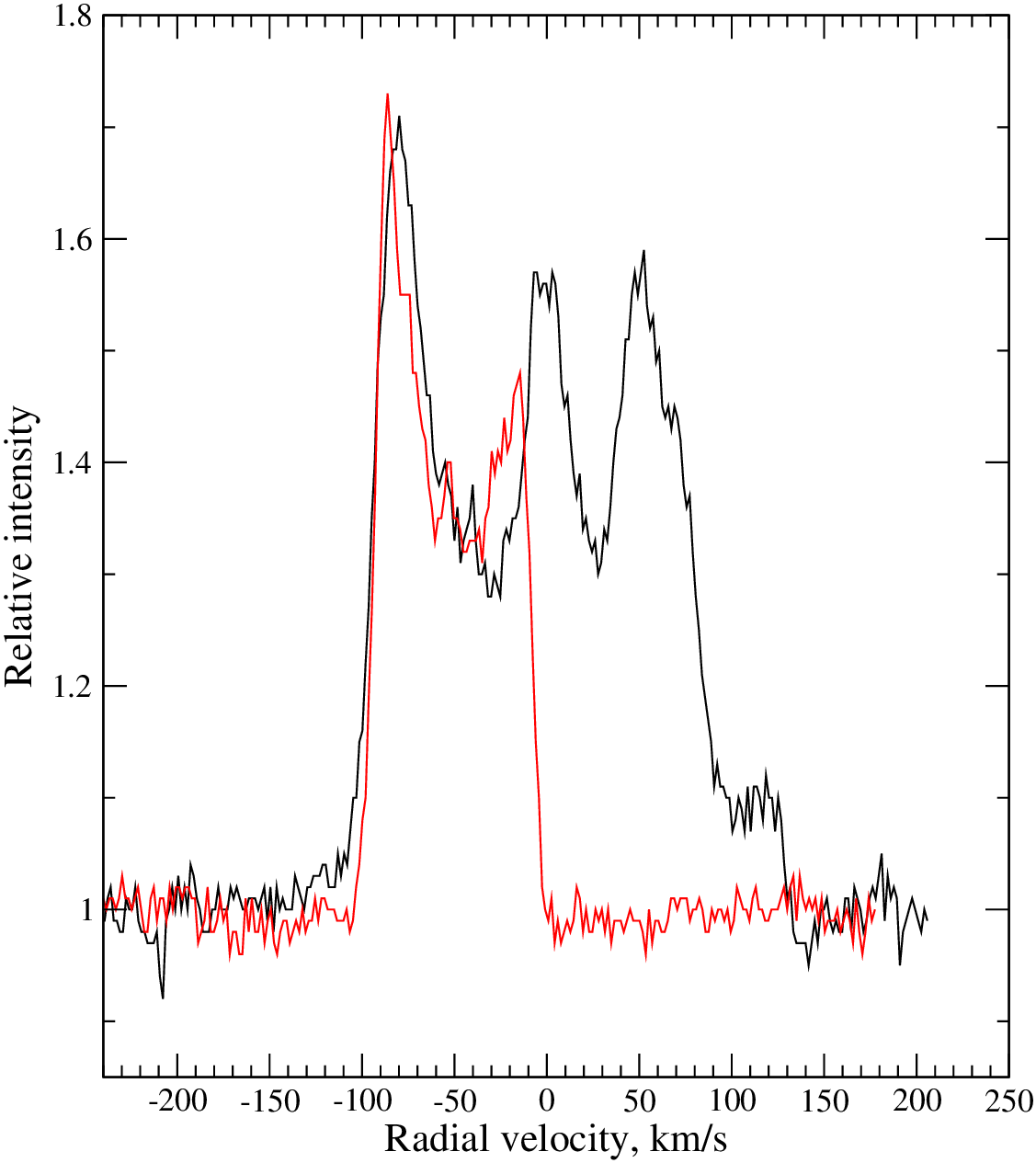}  
\caption{Oxygen triplet O\,I~7773\,\AA{} (black line) in the CI\,Cam, spectrum obtained
on July~5,~2022. The red line is the profile of the two-peaked emission of Fe\,II~5432\,\AA{}
in the same spectrum.}
\label{O7773}
\end{figure}

Another emission feature was recorded in the CI\,Cam spectra --  the O\,I~7773\,\AA{} 
triplet, which is shown in Fig.\,9. An additionally plotted profile  of the two-peaked 
Fe\,II~5432\,\AA{} emission from the same spectrum emphasizes the  coincidence of the 
position of the short-wavelength wing of the left component of the  triplet and the 
Fe\,II~5432\,\AA{} emission. The other two components of the triplet in the figure 
have a blend. The emission of the IR oxygen triplet is a peculiarity of  the spectra 
of hot supergiants. In our studies of supergiants with gas-and-dust  envelopes, we 
previously detected triplet emission in the spectra of the B[e] star in the  
system of the source IRAS\,004770+6429 [24] and in the spectra of the B-supergiant 
LS\,III$+52\degr 24$ (see [25], Fig.\,6). This unstable post-AGB star, associated
with the IR source IRAS\,22023+5249, has a peculiar optical spectrum with a record high  
H$\alpha$ emission intensity and H$\beta$ with signs of winds of up to 290 km/s.

\subsection{Interstellar details}

The complex absorption-emission profiles of the Na\,I doublet remain virtually 
unchanged: measurements of the positions of their components in the available 
spectra  allow the use of average values.

\begin{table}
\medskip
\caption{
Table\,2. Equivalent widths W$_{\lambda}$(DIBs) averaged over CI\,Cam spectra.
The last column shows the corresponding color excess values $E(B-V$)
using calibrations [27].}
\begin{tabular}{c  c  c  }
\hline
 $\lambda$, \AA{}&\hspace{0.1cm}W$_{\lambda}$(DIBs), m\AA{}\hspace{0.2cm} &$E(B-V$),\,mag\\
\hline
5705.20 & 54$\pm$4 & 0.5   \\  [-5pt]
5780.48 &376$\pm$14& 0.7   \\  [-5pt]
5797.06 &126$\pm$12& 0.8   \\  [-5pt]
5849.81 & 38$\pm$6 & 0.5   \\  [-5pt]
6195.98 & 51$\pm$4 & 0.85  \\  [-5pt]
6203.05 &121$\pm$20& 1.0   \\  [-5pt]
6379.32 & 76$\pm$5&0.65    \\  [-5pt]
6613.62 &188$\pm$3 & 0.75  \\  [-5pt]
6660.71 & 51$\pm$4 & 0.9   \\
\hline
\end{tabular}
\label{EW_DIBs}
\end{table}

Average radial velocity for the  permanently present components of Na\,I doublet lines: 
Vr(NaI)=$-7.1\pm 0.5$, $-35.8\pm 0.3$ and  $-77.7\pm 0.8$\,km/s, obtained by averaging
measurements over 12 spectra. In Fig.\,10, both multicomponents Na\,I~D~lines are 
shown in the spectrum obtained on October~2021. In addition to the main components
(circumstellar emission ``1'' and interstellar absorptions ``2'', ``3''),
these profiles also contain a telluric emission ``4''. The dashed vertical lines in 
the figure indicate the average position of the two interstellar components (Vr(KI)=$-7.0$ and
 $-36.7$\,km/s) of the  line K\,I~7696\,\AA{}. The average velocity  Vr(DIBs) in  
11 spectra, Vr(DIBs)=$-7.04\pm 0.21$\,km/s, coincides with the velocity
of the long-wave components of the Na\,I and K\,I lines profiles. Systemic velocity 
of CI\,Cam  V(LSR)$\approx$50\,km/s  and the presence of short-wave interstellar
absorption in the spectrum indicate a significant distance of CI\,Cam in the Perseus arm, 
according to the velocity pattern in the Galaxy [26], which is consistent 
with the parallax of the star $\pi$=0.2101\,mas from the Gaia~DR3 catalog.

By measuring the equivalent widths of the DIBs in the  spectra and using
calibrations [27], we appreciated the excess color (see Table\,2). Its average
value over 9 DIBs bands $E(B-V)=0.74\pm 0.06^m$. Using the standard ratio value
$A_v/E(B-V)$=3.2, we obtain an estimate of an interstellar extinction  $A_v=2.4^m$.
Authors [8], by analyzing absorption estimates obtained from SED modeling, indicate
total extinction  values $A_v=3.71^m$  and higher, up to $4^m$,
which leads to greater distance CI\,Cam, d$>4$\,kpc.

\begin{figure}[ht!]
\includegraphics[angle=0,width=0.45\textwidth,bb=10 70 550 680,clip]{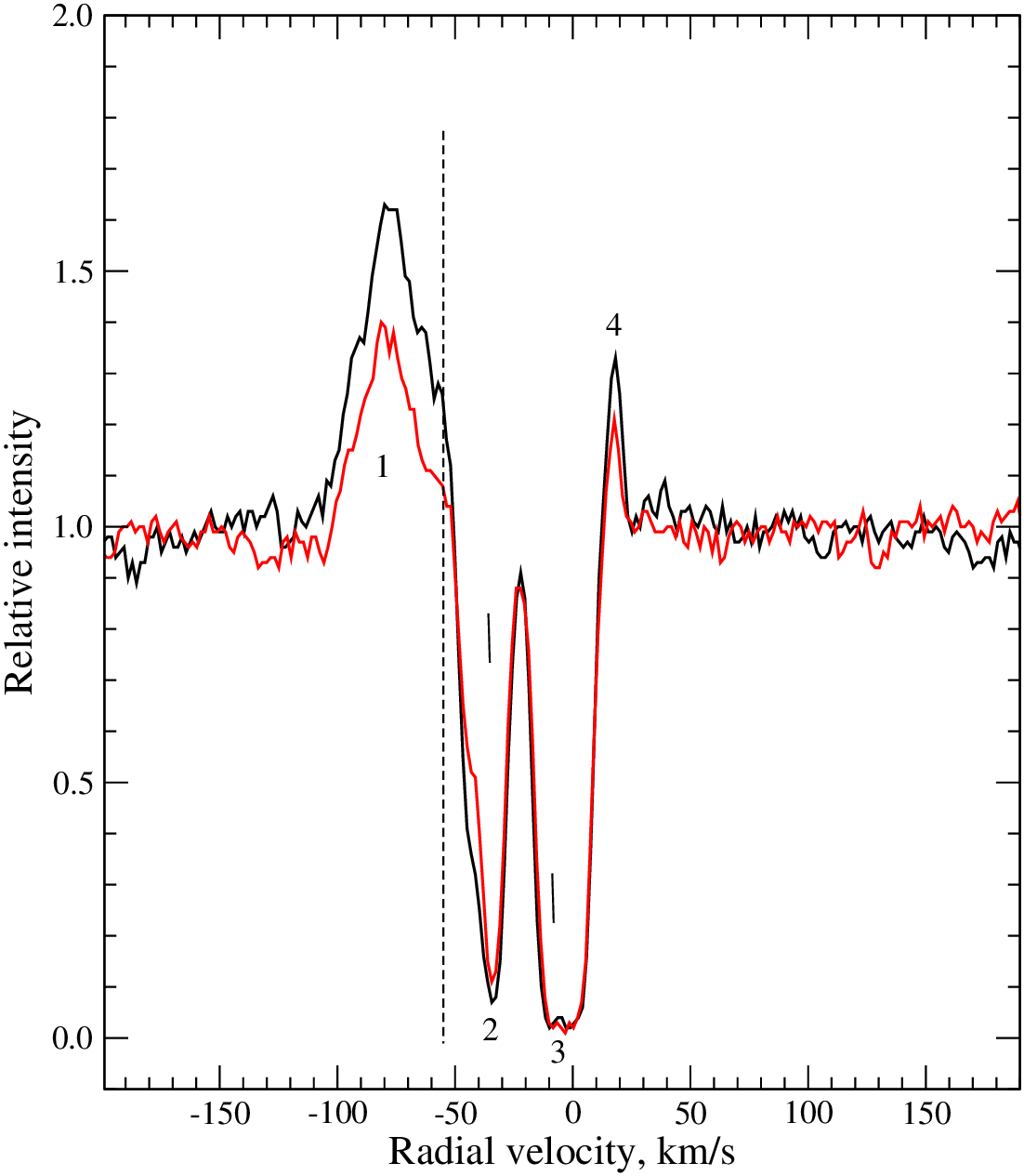}
\caption{Line profiles of Na\,I~5889\,\AA{}  (black line) and Na\,I~5895\,\AA{} (red)
in the CI\,Cam spectrum obtained on October~21,~2021. Short vertical lines indicate
the average position of the two interstellar components, Vr(KI)=$-7.0$\,km/s and
$-36.7$\,km/s, line K\,I~7696\,\AA{} }
\label{NaD}
\end{figure}

\section{Discussion of results}

The group of [Be] stars is heterogeneous, this is well illustrated by the features 
of the observed emission profiles in  [28] as well  examples of model profiles
 of specific lines [7]. The  optical spectra of CI\,Cam contain double-peaked and 
single-peaked emissions from metals as well as single-peaked emissions
from the H\,I and He\,I lines. From the list of main line categories [7], typical 
for the spectra of stars with the B[e] phenomenon, there are no normal profiles of
the P\,Cyg type. At present, it is necessary to confirm the presence of absorptions
arising in the stellar atmosphere, which we found in the spectra of 2022. Among
the forbidden lines, those of low-excitation ions, formed in the outer zone of low 
density, dominate. High excitation lines like [S\,III] are rarely observed. 
In particular, this emission is absent from our CI\,Cam spectra.  However, a pair 
of single-peaked high-excitation [O\,III] emissions at 4959 and 5007\,\AA{} is 
constantly present.
Their intensity is low, only 2--3\% above the level of the local continuum, however, 
their position is stable and the average radial velocity for these forbidden emissions
for each date, indicated in the 5-th column of the Table\,1  is no different from 
the general picture of velocities.

It is also necessary to mention  on the features of the  He\,II~4686\,\AA{} emission, 
whose behaviour  was used [29, 30] to study radial velocity variability in the system 
and determine the orbital period. Using data obtained from various moderate and high 
resolution spectrographs, these authors identified a high amplitude (up to 500\,km/s) 
velocity variability and determined an orbital period of about 19~days. However,
having a homogeneous sample of spectra, we cannot confirm this result and can only 
talk about the rare appearance of emission near the wavelength 4686\,\AA{}. 
Observational data from other authors also do not allow us to identify this 
feature with the high excitation line of He\,II~4686\,\AA{}. Bartlett et al. [3], 
having obtained two high-resolution spectra in October~2016, also did not find 
the presence of the He\,II~4686\,\AA{} feature and highlighted the unreliability 
of using this line to determine the orbital period. Let us also recall the results of
the paper [28], the authors of which, citing characteristic emissions for a sample 
of B[e] stars, emphasize the absence in the spectrum of CI\,Cam of the most
typical emissions for objects of this type: [Ca\,II]~7291 and 7324\,\AA{}, as well 
as [O\,I]~6300 and 6363\,\AA{}. In a later publication [31] the absence of 
He\,II~4686\,\AA{} emission was also noted.

We have already mentioned above the object MWC\,17, which has similar features 
of the optical spectrum.  As shown based on direct H$\alpha$  images authors
[32], both stars, MWC\,17 and CI\,Cam, stand out in the sample of stars 
with the B[e] phenomenon in the absence of extended H$\alpha$-shells. On the 
other hand,  the combination of a number of observed features in  the spectrum of 
CI\,Cam makes this object similar to another supergiant MWC\,137. Long-term study 
allows us to finally classify MWC\,137 as a supergiant with the B[e] phenomenon 
(see details and links in  [33]). In addition to its peculiar optical spectrum, 
this object is also interesting because it is a member of a stellar group and is 
surrounded by an extended H$\alpha$~Sh2-266 nebula.  In the context of our paper, 
the B[e] supergiant MWC\,137 is useful for comparing its properties with those of 
CI\,Cam. In the optical spectrum of MWC\,137 the profile H$\alpha$ has similar 
features to CI\,Cam, except for its extreme width; Fe\,II emissions are 
double-peaked; as in the spectrum of CI\,Cam. The spectra of MWC\,137 lack the 
typical disk emissions [O\,I]~5577\,\AA{}, [Ca\,II]~7291 and 7324\,\AA{}, as well 
as photospheric absorptions. This similarity in the details of the optical
spectrum allows us to classify CI\,Cam as a B[e] supergiant

Analysis of the set of radial velocity measurements allows us to speak about 
the stability of the kinematic state in the CI\,Cam system. Based on measurements 
of the radial velocity from the positions of numerous ``rectangular'' emissions 
in the spectra of the star, a weak variability of the radial velocity was found 
in the interval Vr(emis-d)= $-50.8\div -55.7\pm 0.2$\,km/s. The radial velocity 
corresponding to the position of the forbidden line [N\,II]~5754\,\AA{} is stable
throughout all observations,  Vr=$-55.4\pm 0.6$\,km/s, and we accepted it as systemic. 
Moreover, a comparison of the profiles of numerous ``rectangular'' emissions, 
including [N\,II]~5754\,\AA{} in the spectra of various nights after the 1998 outburst,
indicates a systematic change in the intensity of the profile peaks.

\section{Conclusions}

Below we list our main results and conclusions: 
\begin{itemize}
\item{} for all dates of CI\,Cam observations, ``rectangular'' profiles of 
two-peaked forbidden and permitted emissions of metal ions were recorded. 
The intensities of their blue-shifted  peaks are higher than red-shifted
peaks for all observational dates, with the exception of  the spectrum of 
January~13,~2011 with an inverse V/R ratio and the spectrum of November~18,~2008 
with equal peak intensities;
\item{} a significant decrease in the intensities of all emissions with 
``rectangular'' profiles was found with increasing distance in time from the 1998 outburst;
\item{} the H$\alpha$ is not completely symmetric and contains additional  variable parts. 
At a constant position of the line profile as a whole, changes in the intensity
of the profile top and additional details are observed;
\item{} the kinematic state of the CI\,Cam system is stable, average emission
velocity for all observational dates Vr(aver)= $-53.1\pm 0.5$\,km/s. The half-amplitude 
of the variations (root mean square deviation) is $\Delta$Vr=2.5\,km/s;
\item{} the behavior of the forbidden line [N\,II]\,5754\,\AA{}  does not differ
from those of other emissions: the radial velocity corresponding to its position 
is stable in our observations after 2002 and is accepted by us as systemic, 
Vr=$-55.4\pm 0.6$\,km/s;
\item{}  a rarely observed emission feature was discovered in the spectra of CI\,Cam --- 
the emissive nature of the components of the O\,I~7773\,\AA{}  triplet;
\item{} the conclusion about the absence of emissions of [O\,I]~6300 and 6363\,\AA{}, 
[Ca\,II]~7291 and 7324\,\AA{} was confirmed. The intensity and position of the emission
near 4686\,\AA{} vary significantly. The significant intensity of this line, about 16\% 
of the local continuum, was recorded only in the spectrum obtained in September~2015. 
The currently available high-resolution spectra do not make it possible to resolve the issue
of the reliability of of this feature for estimating the orbital period of the star and the 
localization of this  emission in the CI\,Cam system. The need to increase  the
volume of high-quality spectra is obvious;
\item{} by measuring the equivalent widths of the DIBs and using published calibrations,
 we estimated the color excess $E(B-V)$ due to interstellar extinction. Its
average value over 9 DIBs bands is $E(B-V)=0.74\pm 0.06^m$.   Using the standard ratio
$A_v/E(B-V)$=3.2 we obtain an estimate of interstellar extinction $A_v=2.4\pm 0.2$;
\item{} in the spectra of 2022 and 2023, photospheric absorptions of Fe\,III, N\,II,
Ti\,II and S\,II were identified,  which made it possible to obtain average radial velocity
 with a good accuracy,  Vr(abs-2022)=$-32.4\pm 0.5$\,km/s (5 lines) and
Vr(abs-2023)=$-54.64\pm 1.0$\,km/s  (9 lines), and gives grounds for the conclusion about
the presence of absorptions and the variability of their positions. In the spectrum of
December~8,~2019, we also found 5 absorptions; the average velocity for this date
was determined with lower accuracy, Vr(Vr-2019)=$‒55.6\pm 2.0$\,km/s. The 2021 spectrum
contains only two N\,II absorptions.
\end{itemize}

In general, we can talk about the  stability of the kinematic picture of the CI\,Cam
system in a sleeping  state; we have not recorded any dramatic changes over
the course of 20~years. But, at the same time, the CI\,Cam spectra exhibit essential
variability  in the intensity of emissions of all types, which may be a manifestation
of the inhomogeneity of physical conditions in the circumstellar envelope. We consider
an important result of the work to be the discovery for a number of observational 
moments of photospheric-type absorptions with variability of their average position.
This result also requires the accumulation of high-resolution spectral
data. It is clear that the CI\,Cam system needs continued observation and a deeper 
study.

\section*{Funding}

Observations with the 6-meter telescope of the Special Astrophysical Observatory 
of the Russian Academy of Sciences were supported by the Ministry of Science and 
Higher Education of the Russian Federation.

We thank the Russian Science Foundation for financial support (grant no.\,22--12--00069).

The work used information from the astronomical databases SIMBAD, VALD, SAO/NASA ADS, and Gaia DR3.

\section*{Conflict of interests}

The authors of this work declare that they have no conflicts
of interest.

\end{document}